\documentclass[twocolumn,twocolappendix]{aastex631}

\usepackage{upgreek}
\usepackage{amsmath,amssymb,amsfonts,bm}
\usepackage{epstopdf}
\usepackage{epsfig}
\usepackage{natbib,caption2}
\usepackage{graphicx}   % Including figure files
\usepackage{float}
\usepackage{longtable}
\usepackage{graphics}
\usepackage{hyperref}
\usepackage{color}
\usepackage{calc}
\usepackage{threeparttable}
\usepackage{ulem}
\usepackage{booktabs}%提供命令\toprule、\midrule、\bottomrule
\newcommand \beq{\begin{equation}}
\newcommand \eeq{\end{equation}}
\newcommand \bey{\begin{eqnarray}}
\newcommand \eey{\end{eqnarray}}

\newcommand \kms{\,{\rm km \, s}^{-1}}

\newcommand{\gsim}{\lower.5ex\hbox{$\; \buildrel > \over \sim \;$}}
\newcommand{\lsim}{\lower.5ex\hbox{$\; \buildrel < \over \sim \;$}}

\newcommand{\hb}{\hbox{H$\beta$}}

\newcommand{\oiii}{\hbox{[O\,{\sc iii}]}}

\newcommand{\heii}{\hbox{He\,{\sc ii}}}

\newcommand{\Oiiib}{[O\,{\sc iii}]\,$\lambda$5007}
\newcommand{\Oiiiab}{[O\,{\sc iii}]\,$\lambda\lambda$4959,5007}

\newcommand  \siiv  {\ifmmode {\rm Si}\, {\sc iv}\ \else Si\,{\sc iv}\fi}
\newcommand  \SIIV  {\ifmmode {\rm Si}\,{\sc iv}\,\lambda1399 \else Si\,{\sc iv}\,$\lambda1399$\fi}
\newcommand  \civ  {\ifmmode {\rm C}\, {\textsc iv}\ \else C\,{\sc iv}\fi}
\newcommand  \CIV  {\ifmmode {\rm C}\,{\sc iv}\,\lambda1549 \else C\,{\sc iv}\,$\lambda1549$\fi}

\newcommand  \aliii  {\ifmmode {\rm Al}{\textsc{iii}} \else Al\,{\sc iii}\fi}
\newcommand  \ALIII  {\ifmmode {\rm Al}\,{\sc iii]}\,\lambda1854 \else C\,{\sc iii]}\,$\lambda1854$\fi}
\newcommand  \feii {\ifmmode {\rm Fe}\,{\textsc{ii}}\, \else Fe\,{\sc ii}\fi}

\newcommand  \MGII  {\ifmmode {\rm Mg}\,{\sc ii}\,\lambda2798 \else Mg\,{\sc ii}\,$\lambda2798$\fi}

\newcommand{\lya}{Ly~$\alpha$}

\def\kms{$\rm km\,s^{-1}$}

\def\ergs{${\rm erg\,s^{-1}}$}

\shortauthors{Chen et al.}
\begin{document}

\title{Narrow \oiii\ emission lines as a potential proxy for the evolutionary stage of quasars}

\correspondingauthor{Zhe-Geng Chen}
\email{zhegengc@126.com}

\author{Zhi-fu Chen}
\affiliation{School of Mathematics and Physics, Guangxi Minzu University, Nanning 530006, People's Republic of China}

\author{Zhe-Geng Chen}
\affiliation{Laboratory for Relativistic Astrophysics, Physical Science and Technology College, Guangxi University, Nanning 530004, People's Republic of China; zhegengc@126.com or zhichenfu@126.com}
%\affiliation{School of Mathematics and Physics, Guangxi Minzu University, Nanning 530006, People's Republic of China}

\author{Xing-long Peng}
\affiliation{School of Mathematics and Physics, Guangxi Minzu University, Nanning 530006, People's Republic of China}

\author{Wei-rong Huang}
\affiliation{School of Mathematics and Physics, Guangxi Minzu University, Nanning 530006, People's Republic of China}

%\collaboration{6}{(AAS Journals Data Editors)}

%% Note that the \and command from previous versions of AASTeX is now
%% depreciated in this version as it is no longer necessary. AASTeX
%% automatically takes care of all commas and "and"s between authors names.

%% AASTeX 6.31 has the new \collaboration and \nocollaboration commands to
%% provide the collaboration states of a group of authors. These commands
%% can be used either before or after the list of corresponding authors. The
%% argument for \collaboration is the collaboration identifier. Authors are
%% encouraged to surround collaboration identifiers with ()s. The
%% \nocollaboration command takes no argument and exists to indicate that
%% the nearby authors are not part of surrounding collaborations.

%% Mark off the abstract in the ``abstract'' environment.
\begin{abstract}
Radio spectral shape of quasars can provide insight into the ages of quasars. We have compiled data for 1804 quasars with $z\lesssim1$ from the Sloan Digital Sky Survey (SDSS). Additionally, these quasars were also mapped by the Low-Frequency Array at 144 MHz and the Very Large Array Sky Survey at 3000 MHz. The radio spectral index, designated as $\alpha^{\rm 144}_{\rm 3000}$ (with $S(\nu)\propto\nu^\alpha$), is analyzed between 144 MHz and 3000 MHz as a proxy for the ages of quasars. We measure the \Oiiib\ emission line in the SDSS spectra. A strong correlation was found between the equivalent width of the core component of the \Oiiib\ emission line and $\alpha^{\rm 144}_{\rm 3000}$. This relationship suggests that the core component of the \Oiiib\ emission line could potentially serve as a surrogate for the evolutionary stage of a quasar. The quasars at an early stage of evolutions tend to show weaker \Oiiib\ emission, while older quasars exhibit stronger \Oiiib\ emission.
\end{abstract}


\section{Introduction}
t is a widely accepted fact that supermassive black holes (SMBHs) reside at the center of host galaxies. These SMBHs power active galactic nuclei (AGN) through the accretion of surrounding material. This process releases large amounts of energy that, in the form of feedback, influences the physical conditions and dynamics of gas and dust within the host galaxies. Consequently, it is postulated that SMBHs and their host galaxies co-evolve and mutually regulate each other's development \citep[e.g.][]{2001ApJ...547..140M,2003ApJ...589L..21M,2004ApJ...604L..89H,2009MNRAS.397.1705G,2013ARA&A..51..511K,2014ARA&A..52..589H,2017NewAR..79...59X,2018MNRAS.481.3278R,2022RAA....22a5010L}. In order to thoroughly investigate the evolution of SMBHs and/or host galaxies, we must address a fundamental question: how can we characterize the corresponding objects that exist at different stages of evolution?

Unlike galaxies and stars, determining the ages of quasars is a challenging task. Various methods used to estimate quasar ages often yield a large uncertainty spanning millions of years \cite[e.g.,][]{2004cbhg.symp..169M,2008MNRAS.391.1457K,2014MNRAS.442.3443D,2017NatAs...1..596M}. The properties of absorption lines serve as one common method used to characterize the evolutionary stages of quasars \citep[e.g.,][]{1982MNRAS.198...91C,1988ApJ...327..570B,1992ApJ...399L..15B,1996ApJ...466...46G,2000ApJS..130...67S,2012MNRAS.427.1209C,2012ApJ...748..131S,2015ApJ...806..142Z,2016ApJ...824..106W,2019MNRAS.484.3897K,2022NatAs...6..339C,2022SciA....8.3291H,2024ApJ...963....3P}.
Low-ionization associated absorption lines (AALs) typically indicate quasars in early stages of evolution, while quasars without AALs or those with only high-ionization AALs are considered to be in later stages of evolution. The proximity effect, which describes a decrease in the number density of \lya\ or \heii\ forest absorption lines observed in quasar sightlines \citep[e.g.,][]{1982MNRAS.198...91C,1988ApJ...327..570B,2019MNRAS.484.3897K,2019ApJ...883..123Z}, is yet another method employed for aging quasars. According to the proximity effect, a small proximity zone suggests a younger quasar. However, using absorption lines to determine quasar ages leaves some questions unanswered. The detection and properties of AALs in quasars could be affected by viewing angle \citep[e.g.,][]{1995PASP..107..803U,2012ASPC..460...47H}. Furthermore, the proximity effect has its limitations. Neither the \lya\ nor the \heii\ forest absorption lines can be applied to UV-optical spectra of quasars with low redshifts (e.g., $z<1.5$).

The powerful UV radiation from quasars produces ionization zones within or around host galaxies, and is the primary mechanism behind the proximity effect observed in forest absorption lines. Additionally, it can alter the physical conditions within narrow-line regions (NLRs). The cumulative ionizing radiation field promotes the growth of NLRs. Therefore, changes in narrow emission lines (NELs) can provide clues to a quasar's age. This is corroborated by \cite{2020ApJ...892..139Z}, who found that young quasars only host weak NELs, while older ones exhibit strong NELs.

According to the unified schemes of AGNs \cite[][]{1995PASP..107..803U}, the NLR is located outside the dust torus and can extend from 100 pc to several kpc \cite[e.g.,][]{2024Galax..12...17H}. The broad-line region (BLR) is located inside the dust torus and is often obscured by it. The position of the NLR means its emission lines are less affected by quasar viewing angles compared to broad emission lines (BELs) that originate in BLRs. What's more, the NLR is compact enough to be illuminated by radiation from the central regions of the quasar. Thus, NELs with high ionizations, such as \Oiiiab, are often observed in quasar spectra and can be detected by ground-based telescopes for objects at redshifts $z\lesssim1$. Consequently, the NLR serves as an excellent location for studying quasar properties. In this paper, we use narrow \oiii emission lines to investigate the growth of NLRs throughout the life cycle of radio quasars.

Throughout this work, we assume a flat $\Lambda$CDM cosmology with $\Omega_m$ = 0.3,  $\Omega_\Lambda$ = 0.7, and $h_0$ = 0.7.

\section{Data samples and spectral analysis} \label{sec:sample}
\subsection{Data samples}
The Sloan Digital Sky Survey (SDSS) provided spectra of 750,414 quasars in their Sixteenth Data Release (DR16Q) \citep[][]{2000AJ....120.1579Y,2020ApJS..250....8L}. The spectra were obtained in wavelength ranges of $\lambda \approx 3800-9200$ {\AA} in the SDSS-I/II \citep[]{2009ApJS..182..543A}, or $\lambda \approx 3600-10500$ {\AA} in the SDSS-III/IV \citep[]{2013AJ....146...32S,2013AJ....145...10D}. To measure the \oiii\ emission lines, we drew from the SDSS DR16Q spectra of quasars with $z\lesssim1.1$, where the $z$ represents the improved system redshifts ($Z_{\rm sys}$) as annotated by \cite{2022ApJS..263...42W}. This results in our parent sample of 114,955 quasars.

The synchrotron radiation stemming from relativistic electrons, which maintain an initial power-law electron energy distribution and continuously move through a magnetic field \citep[e.g.,][]{1962SvA.....6..317K,1964ApJ...140..969K,1971PhT....24i..57P}, generates radio spectra of quasars that follow the relation $S(\nu )\propto \nu^{\rm \alpha}$. Here, $S(\nu)$ represents the observed flux density at frequency $\nu$, and $\alpha$ is the spectral index. During the quasar's evolution, the shape of the radio spectrum can change due to the absorption of synchrotron radiation \citep[e.g.,][and references therein]{2017NatAs...1..596M,2023A&A...674A.198K}. Young quasars typically have smaller-scale jets and their radio spectrum often peaks in the GHz frequency range. As quasars evolve and their jets grow, the peak of the radio spectrum shifts from high to low frequencies. Thus, the shape of a quasar's radio spectrum is considered as a reliable tracer of its life cycle \citep[e.g.,][]{1994A&A...285...27K,2017NatAs...1..596M,2020A&A...638A..29B,2021Galax...9...88M,2023A&A...674A.198K}.

The Low-Frequency Array (LOFAR) \citep[][]{2017A&A...598A.104S} Two-metre Sky Survey (LoTSS) \citep[][]{2013A&A...556A...2V} maps the northern sky with a central frequency of 144 MHz and an angular resolution of $6''$. Its second data release (DR2) includes data for 4,396,228 radio sources \citep[][]{2022A&A...659A...1S}. The Very Large Array Sky Survey (VLASS) \citep[][]{2020PASP..132c5001L} maps the sky at a central frequency of 3000 MHz (2 --- 4 GHz) with an angular resolution of $\sim2.5''$ and includes data for more than 2.6 million radio sources (available at $https://cirada.ca/catalogs$) \citep[][]{2021ApJS..255...30G}. To trace the radio spectral shape of SDSS quasars, we cross-referenced our parent sample (114,955 quasars) with both VLASS and LoTSS DR2 data within a radius range of $7''$. As a result, we obtained data for a sample of 2288 quasars from whom we determined the radio spectral index between 144 MHz and 3000 MHz ($\alpha^{\rm 144}_{\rm 3000}$). Here the $\alpha^{\rm 144}_{\rm 3000}$ is computed via
\begin{equation}\label{eq:index}
  \alpha^{\rm 144}_{\rm 3000} = \frac{Log S(\nu)^{\rm VLASS} - Log S(\nu)^{\rm LoTSS}}{Log \nu_{\rm VLASS} - Log \nu_{\rm LoTSS}},
\end{equation}
where the $S(\nu)^{\rm VLASS}$ and $S(\nu)^{\rm LoTSS}$ are the flux densities at 3000 MHz and 144 MHz, respectively. This index characterizes the spectral shape of the radio spectrum for these quasars.

\subsection{Spectral analysis}
Our methods for fitting the SDSS spectra to minimize $\chi^2$ are modeled after earlier works \citep[e.g.,][]{2018ApJS..234...16C,2019ApJS..244...36C,2023ApJS..264...52H}. Initially, we correct the quasar spectra for Galactic extinction, employing the reddening maps put forth in \cite{2011ApJ...737..103S} and using the Milky Way extinction curve from \cite{1989ApJ...345..245C}. Subsequently, for each spectrum, we fit a local power-law continuum ($f_{\rm \lambda}=A\lambda^{\rm \alpha}$) and the iron template \citep[][]{2001ApJS..134....1V,2004A&A...417..515V} to the data within the [4400,4800] \AA\ and [5100,5550] \AA\ ranges. We then subtract the resulting continuum+iron fits from the spectrum. The residual spectra are subsequently used to fit the \hb\ and \oiii\ emission lines. We represent the emission lines with multiple Gaussian functions. Broad \hb\ is depicted by three Gaussian functions each with $\rm FWHM>1200$ \kms, while narrow \hb\ is represented by a single Gaussian function. Each line of the \Oiiiab\ doublet is depicted by two Gaussian functions: one for the core (narrow) component with $\rm FWHM<1200$ \kms\ and another for the blue wing (broad) component with $\rm FWHM<2500$ \kms. We force all narrow emission lines to have the same $\rm FWHM$ and velocity offsets from the quasar systemic redshifts. The $\rm FWHM$ and velocity offsets of the wing component of the \oiii\ are also constrained to the same values, and the \Oiiiab\ doublet is forced to maintain a flux ratio of $F(5007)/F(4959)=3$.

\begin{figure}[ht!]
\includegraphics[width=0.48\textwidth]{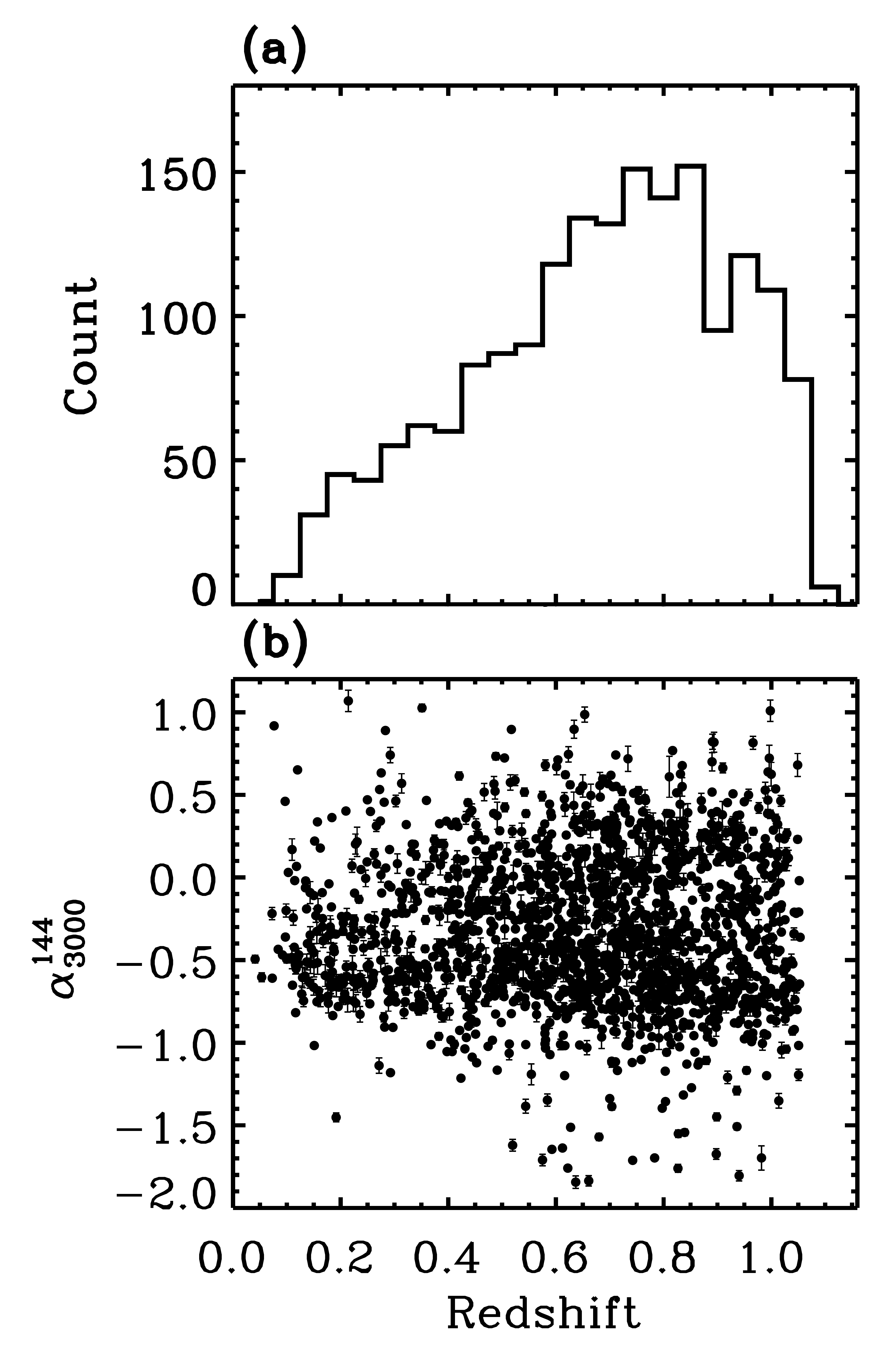}
\caption{(a): The redshift distribution of quasars in our final sample. (b): The relationship between $\alpha^{\rm 144}_{\rm 3000}$ and redshifts. The Spearman's correlation coefficient is $r=-0.008$.}
\label{fig:z}
\end{figure}

We only include measurements where $W_{\rm core}^{\oiii\lambda5007}>4\sigma_{\rm w}$. Here, $W_{\rm core}^{\oiii\lambda5007}$ represents the equivalent width of the core component of \Oiiib, and $\sigma_{\rm w}$ stands for the corresponding uncertainty. Imposing this limit, we end up with a final sample of 1804 quasars, whose redshifts are depicted in Figure \textcolor{blue}{\ref{fig:z}}. We list the measurements and parameters for these quasars in Table \textcolor{blue}{\ref{tab:properties}}.

It is noted from Equation \textcolor{blue}{(\ref{eq:index})} that although the radio spectral index ($\alpha^{\rm 144}_{\rm 3000}$) is computed in the observed frame, the $\alpha^{\rm 144}_{\rm 3000}$ shouldn't be affected by redshifts. Figure \textcolor{blue}{\ref{fig:z}} also shows the distribution of $\alpha^{\rm 144}_{\rm 3000}$ as a function of redshifts, which clearly demonstrates that there is not a significant relationship between the $\alpha^{\rm 144}_{\rm 3000}$ and redshifts.

\begin{table*}[htbp]
\centering
\caption{The properties of the sample}
\tabcolsep 0.7mm \footnotesize%\small%\tiny%
\label{tab:properties}
\begin{tabular}{cccccccccccccccc}
\hline\hline\noalign{\smallskip}
SDSS name & PLATE&MJD&FIBER&$z_{em}$ &  $Log L_{\rm 5100}$  & $Log L_{\rm core}^{\oiii\lambda5007}$ & $W_{\rm core}^{\oiii\lambda5007}$ &$S_{peak}^{LoTSS}$&$S_{peak}^{VLASS}$&$\alpha^{\rm 144}_{\rm 3000}$ \\
&&&&&\ergs&\ergs &\AA &mJy&mJy&  \\
\hline\noalign{\smallskip}
000057.79+294236.4	&	7134	&	56566	&	394	&	0.7595 	&	44.383$\pm$0.078 	&	42.324$\pm$0.017 	&	64.556$\pm$0.590 	&	3.281$\pm$0.223 	&	57.680$\pm$0.387 	&	-0.898 	\\
000101.04+240842.5	&	7666	&	57339	&	82	&	0.9692 	&	44.910$\pm$0.136 	&	42.526$\pm$0.057 	&	43.053$\pm$0.986 	&	17.017$\pm$0.209 	&	202.279$\pm$1.096 	&	-0.769 	\\
000131.63+165413.7	&	6172	&	56269	&	666	&	0.9359 	&	44.935$\pm$0.054 	&	42.865$\pm$0.018 	&	65.429$\pm$0.438 	&	2.603$\pm$0.183 	&	150.350$\pm$0.334 	&	-1.290 	\\
000132.36+211336.2	&	7595	&	56957	&	190	&	0.4390 	&	44.442$\pm$0.024 	&	41.777$\pm$0.020 	&	28.247$\pm$0.131 	&	152.235$\pm$0.290 	&	643.666$\pm$0.962 	&	-0.429 	\\
000141.79+304114.9	&	7749	&	58073	&	266	&	0.9214 	&	44.288$\pm$0.221 	&	42.105$\pm$0.041 	&	47.832$\pm$1.144 	&	4.332$\pm$0.214 	&	3.387$\pm$0.165 	&	0.127 	\\
000316.31+245938.3	&	7666	&	57339	&	58	&	0.6729 	&	44.417$\pm$0.050 	&	41.589$\pm$0.084 	&	29.566$\pm$0.351 	&	60.591$\pm$0.150 	&	82.691$\pm$0.402 	&	-0.057 	\\
000334.90+200942.9	&	7595	&	56957	&	48	&	0.1098 	&	42.924$\pm$0.043 	&	39.823$\pm$0.065 	&	18.086$\pm$0.183 	&	1.565$\pm$0.167 	&	1.077$\pm$0.126 	&	0.169 	\\
000422.96+222127.1	&	7595	&	56957	&	941	&	0.6112 	&	44.493$\pm$0.044 	&	41.890$\pm$0.027 	&	26.830$\pm$0.224 	&	6.596$\pm$0.194 	&	19.815$\pm$0.087 	&	-0.316 	\\
000446.05+222702.6	&	6879	&	56539	&	184	&	0.7077 	&	43.976$\pm$0.108 	&	42.057$\pm$0.013 	&	76.992$\pm$0.580 	&	9.665$\pm$0.189 	&	187.287$\pm$0.366 	&	-0.930 	\\
000452.27+335501.7	&	7748	&	58396	&	712	&	0.4598 	&	44.333$\pm$0.022 	&	41.707$\pm$0.029 	&	38.314$\pm$0.169 	&	0.947$\pm$0.134 	&	3.941$\pm$0.203 	&	-0.424 	\\
\noalign{\smallskip}
\hline\hline\noalign{\smallskip}
\end{tabular}
\begin{flushleft}
(This table is available in its entirety in machine-readable form.)
\end{flushleft}
\end{table*}

\section{Discussions} \label{sec:DISCUS}
The sample we used in our study consists only of type-I quasars. The \oiii\ emission lines can be excited by radiation from the central regions of the quasar (photoionization) as well as the shock from outflows/winds. This is the primary reason why researchers typically use two Gaussian functions to model each line of the \oiii\ doublet - one Gaussian function attempts to represent the photoionization component (core component), while the other represents the outflow component (wing component). The core component has been found to be useful in investigations concerning the physical conditions and kinematics of NLRs or host galaxy \citep[e.g.,][]{2005ApJ...627..721G,2008ApJ...680..926K,2011ApJ...737...71Z,2013MNRAS.433..622M,2013ARA&A..51..511K,2017ApJ...839..120W,2019ApJ...878..101S,2022ApJS..261...23Z,2023ApJ...945...59L,2023ApJ...950...16J}. Consequently, in this study, we use the core component of \Oiiib\ to trace the development of NLRs.

\begin{figure*}[ht!]\centering
\includegraphics[width=0.49\textwidth]{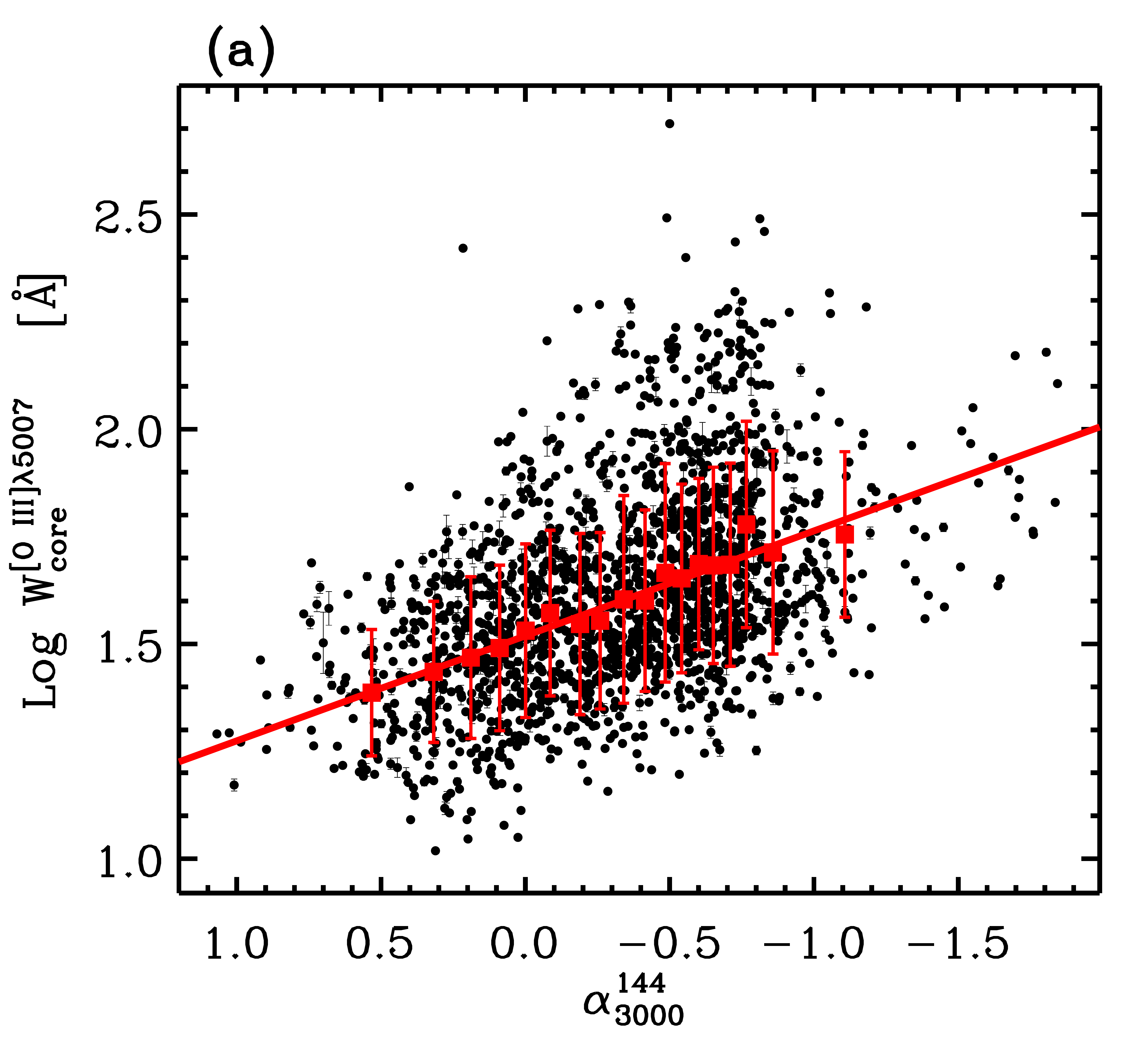}
\includegraphics[width=0.49\textwidth]{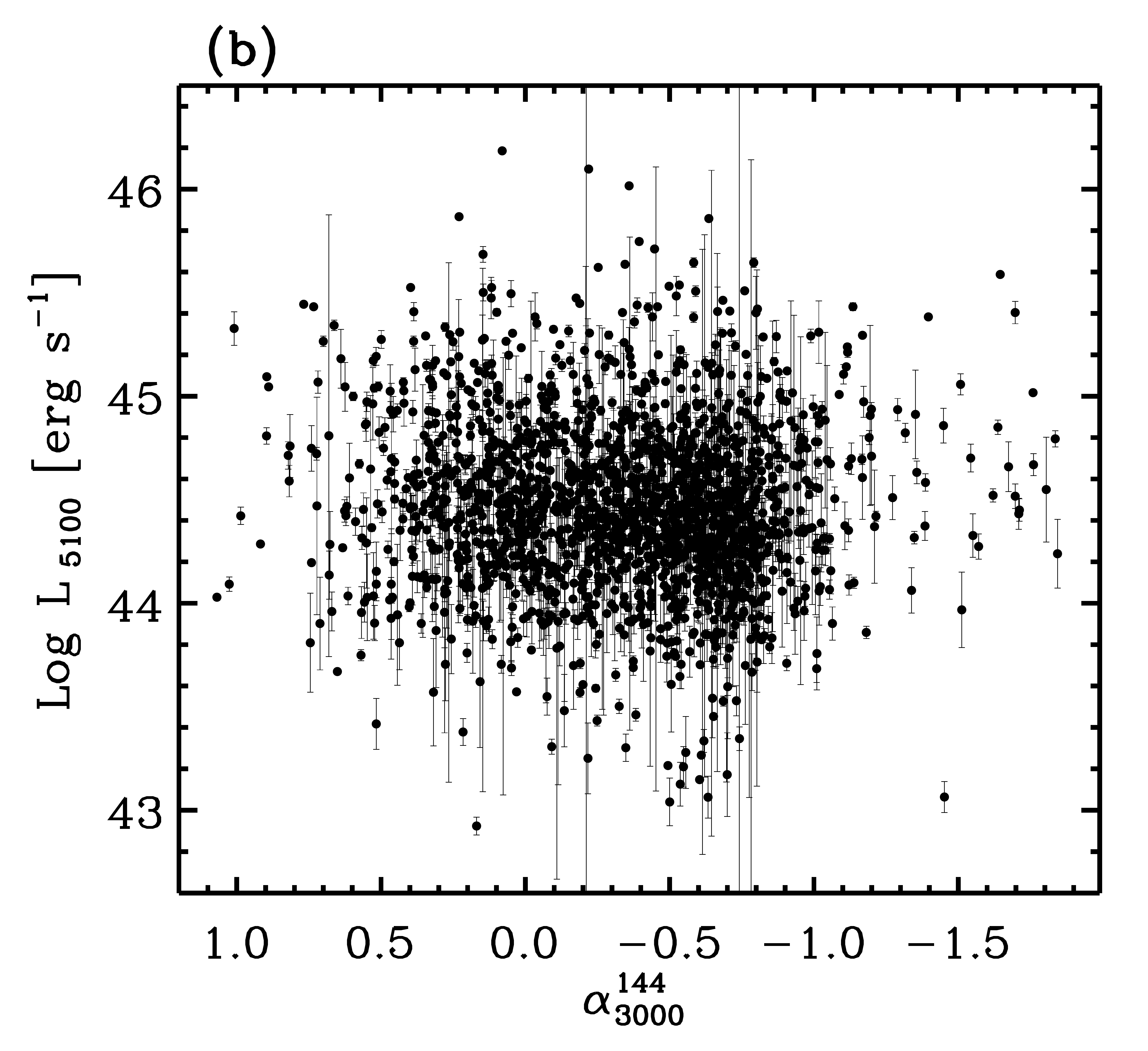}
\includegraphics[width=0.49\textwidth]{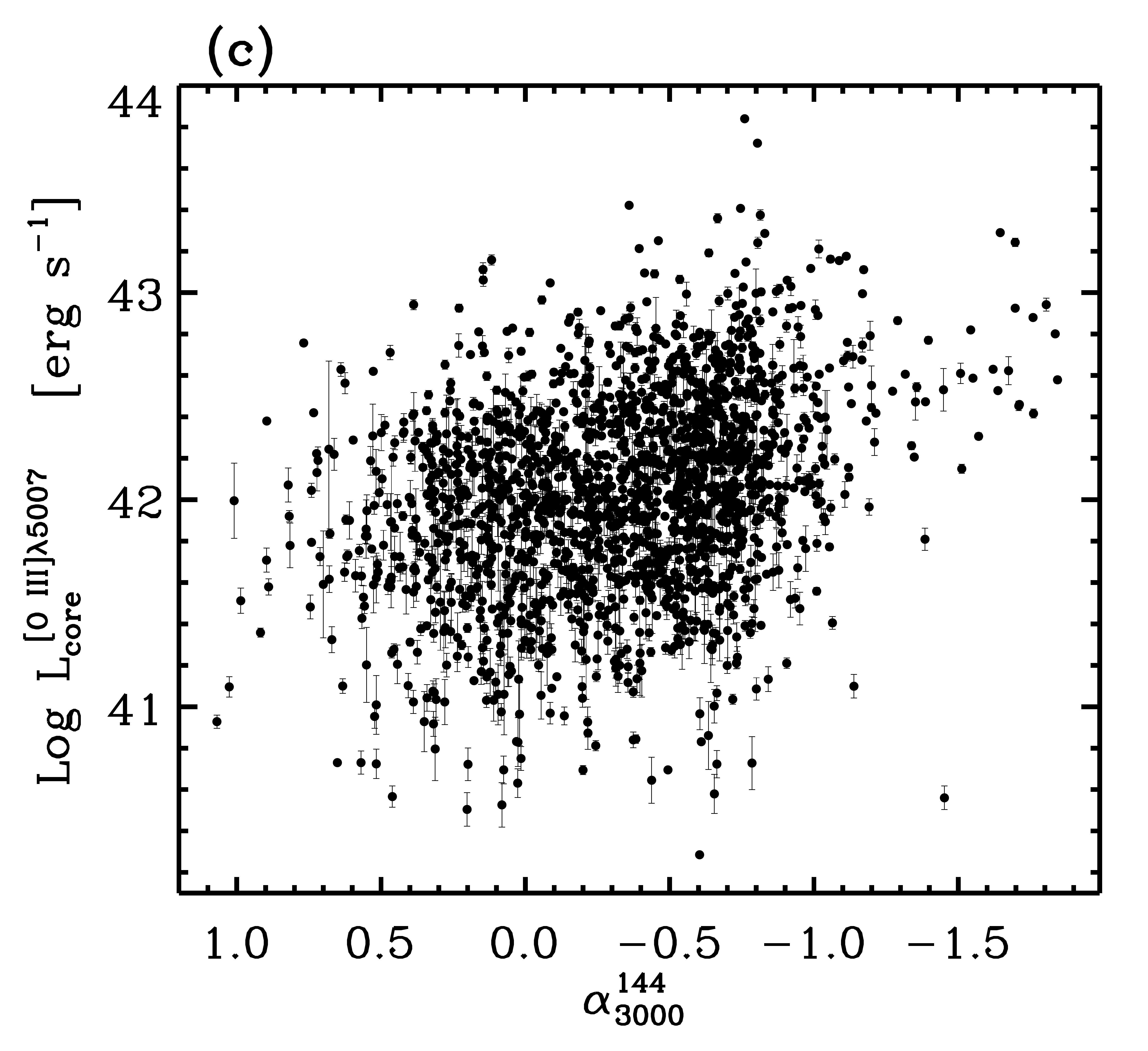}
\caption{(a): The equivalent width of the core component of the \Oiiib\ evolved with radio spectral shape. The large $\alpha^{\rm 144}_{\rm 3000}$ indicates the quasar living in the early evolutionary stage, and the small $\alpha^{\rm 144}_{\rm 3000}$ indicates the evolved quasar. The Spearman's correlation coefficient is $r=-0.489$. Red squares show the median values in distinct $\alpha^{\rm 144}_{\rm 3000}$ bins with the error bars representing standard deviations. Red-solid line is the best linear fitting to the red squares: $Log~W_{\rm core}^{\oiii\lambda5007}\propto (-0.244\pm0.102)\times Log~\alpha^{\rm 144}_{\rm 3000}$. (b): The relationship between continuum luminosity at 5100 \AA\ and $\alpha^{\rm 144}_{\rm 3000}$. The Spearman's correlation coefficient is $r=0.060$. (c): The relationship between \Oiiib\ luminosity of the core component and $\alpha^{\rm 144}_{\rm 3000}$. The Spearman's correlation coefficient is $r=-0.309$.}
\label{fig:alpha_w_LL}
\end{figure*}

The spectral shape of the radio spectrum serves as an effective indicator of a quasar's evolutionary stage. As shown in Figure \textcolor{blue}{\ref{fig:alpha_w_LL}(a)}, there is a notable correlation between the equivalent width of the \Oiiib\ emission line and the radio spectral shape, with a Spearman's correlation coefficient of $r=-0.489$ and a probability of $P<10^{-15}$. BELs are known to exhibit a strong inverse correlation between their equivalent width and the continuum luminosity, a relationship known as the Baldwin effect \citep[][]{1977ApJ...214..679B}. While this effect is less pronounced for NELs \citep[e.g.,][]{2011ApJ...737...71Z,2013ApJ...762...51Z}, it may still significantly influence the relationship between the equivalent width of the \Oiiib\ emission line and $\alpha^{\rm 144}_{\rm 3000}$. To assess the influence of the Baldwin effect, we examine in Figure \textcolor{blue}{\ref{fig:alpha_w_LL}(b)} the distribution of the continuum luminosity at 5100 \AA\ as a function of $\alpha^{\rm 144}_{\rm 3000}$. The data suggests a negligible evolution in the continuum luminosity with respect to the radio spectral shape - an indication that the Baldwin effect likely contributes minimally to the strong correlation between the $W_{\rm core}^{\oiii\lambda5007}$ and $\alpha^{\rm 144}_{\rm 3000}$. The observed tight correlation between the $W_{\rm core}^{\oiii\lambda5007}$ and $\alpha^{\rm 144}_{\rm 3000}$ might be primarily attributed to the growth of NLRs. This correlation is further corroborated by the significant relationship between $L_{\rm core}^{\oiii\lambda5007}$ and $\alpha^{\rm 144}_{\rm 3000}$, as shown in Figure \textcolor{blue}{\ref{fig:alpha_w_LL}(c)}. In essence, younger quasars inherently exhibit weaker \oiii\ emission lines, while older ones host more potent \oiii\ emission lines. These observations align with a study by \cite{2020ApJ...892..139Z}, which reported that narrow emission lines are significantly stronger in older quasars compared to younger ones.

\begin{figure}[ht!]
\includegraphics[width=0.46\textwidth]{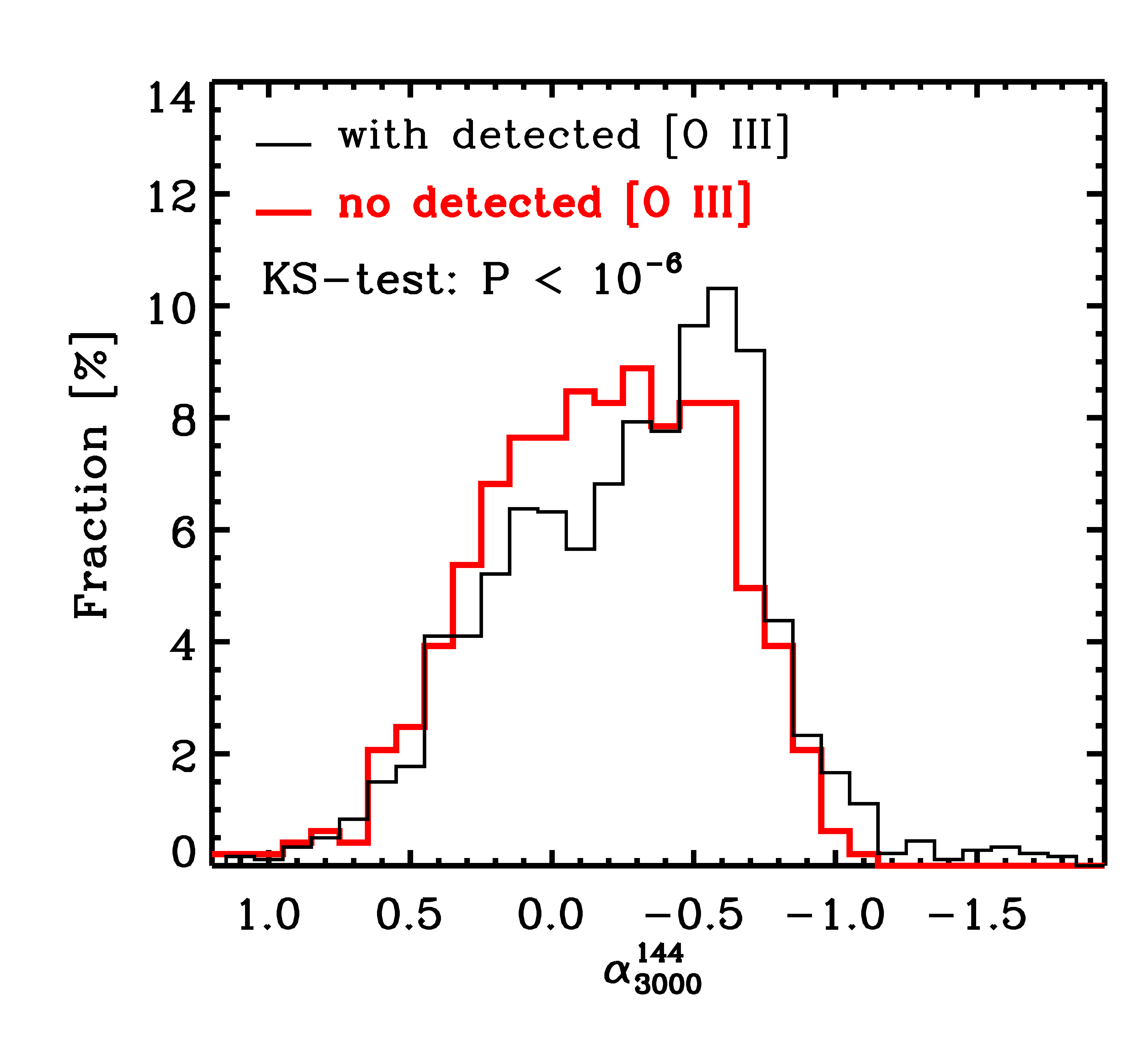}
\caption{The distribution of the $\alpha^{\rm 144}_{\rm 3000}$. The black thin line is for the quasars with detected \oiii\ emission lines, and the red thick line is for the quasars without detected \oiii\ emission lines. The KS-test yields a probability $P<10^{-6}$.}
\label{fig:distr_alpha}
\end{figure}

Out of our parent sample, 484 quasars exhibit \oiii\ emission lines that are too weak or unreliable to be considered ($W_{\rm core}^{\oiii\lambda5007}<4\sigma_{\rm w}$), thus excluded from our final sample. Looking at the bigger picture, the $\alpha^{\rm 144}_{\rm 3000}$ values of these quasars, wherein \oiii\ is undetected, are generally larger compared to those of quasars with detected \oiii\ (refer to Figure \textcolor{blue}{\ref{fig:distr_alpha}}). This observation implies that quasars with undetected \oiii\ might be in an earlier stage of evolution. Consequently, it is reasonable to expect an increase in the correlation between $W_{\rm core}^{\oiii\lambda5007}$ and $\alpha^{\rm 144}_{\rm 3000}$ once reliable measurements become available for the core \Oiiib\ of quasars without detected \oiii.

%The strong correlation between the $W_{\rm core}^{\oiii\lambda5007}$ and $\alpha^{\rm 144}_{\rm 3000}$ suggests that \Oiiib\ could serve as a suitable indicative proxy of the evolutionary stage of quasars. The intensity of narrow \oiii\ emission lines tends to increase as the quasar ages. Meanwhile, \Oiiib\ is the most intense narrow emission line noticeable in the optical spectra of quasars and it can be detected using ground-based telescopes for sources with $z\lesssim1$. As such, the \oiii\ emission line proves to be a valuable addition to the proximity effect of forest absorption lines, a method typically employed in dating high-redshift quasars.

\section{Summaries} \label{sec:summary}
Using the observations from the SDSS, VLASS, and LoTSS, we compiled a sample of 2288 quasars with $z\lesssim1$. In term of the observations from the VLASS and LoTSS, we computed the radio spectral index between 144 MHz and 3000 MHz ($\alpha^{\rm 144}_{\rm 3000}$), which is served as a reliable proxy for quasar ages. Based on the measurements from the SDSS spectra, 1804 quasars have $W_{\rm core}^{\oiii\lambda5007}>4\sigma_{\rm w}$, where the $W_{\rm core}^{\oiii\lambda5007}$ is the equivalent width of the core component of the \Oiiib\ emission line. We find that the $W_{\rm core}^{\oiii\lambda5007}$ is obviously correlated with the $\alpha^{\rm 144}_{\rm 3000}$, which is independent of the Baldwin effect.

The strong correlation between the $W_{\rm core}^{\oiii\lambda5007}$ and $\alpha^{\rm 144}_{\rm 3000}$ suggests that \Oiiib\ could serve as a suitable indicative proxy of the evolutionary stage of quasars. The intensity of narrow \oiii\ emission lines tends to increase as the quasar ages. Meanwhile, \Oiiib\ is the most intense narrow emission line noticeable in the optical spectra of quasars and it can be detected using ground-based telescopes for sources with $z\lesssim1$. As such, the \oiii\ emission line proves to be a valuable addition to the proximity effect of forest absorption lines, a method typically employed in dating high-redshift quasars.

%A strong correlation was found between the equivalent width of the core component of the \Oiiib\ emission line and $\alpha^{\rm 144}_{\rm 3000}$. This relationship suggests that the core component of the \Oiiib\ emission line could potentially serve as a surrogate for the evolutionary stage of a quasar. The quasars at an early stage of evolutions tend to show weaker \Oiiib\ emission, while older quasars exhibit stronger \Oiiib\ emission.

\section*{Acknowledgements}
We deeply thank the anonymous referees for her/his helpful and careful comments. This work is supported by the Guangxi Natural Science Foundation (2024GXNSFDA010069), the National Natural Science Foundation of China (12073007), and the Scientific Research Project of Guangxi University for Nationalities (2018KJQD01).

\bibliography{cG}{}
\bibliographystyle{aasjournal}

%% This command is needed to show the entire author+affiliation list when
%% the collaboration and author truncation commands are used.  It has to
%% go at the end of the manuscript.
%\allauthors

%% Include this line if you are using the \added, \replaced, \deleted
%% commands to see a summary list of all changes at the end of the article.
%\listofchanges

\end{document}